\documentclass[%
 reprint,
 superscriptaddress,
 longbibliography,
 amsmath,amssymb,
 aps,
prb,
]{revtex4-2}

\usepackage{graphicx}
\usepackage{dcolumn}
\usepackage{bm}

\usepackage{xcolor}
\usepackage{soul}
\usepackage{physics}
\usepackage{bbold}

\begin{document}

\preprint{APS/123-QED}


\title{Scalable quantum memory nodes using nuclear spins in Silicon Carbide}
\author{Shravan Kumar Parthasarathy}
\affiliation{Fraunhofer Institute for Integrated Systems and Device Technology  (IISB), Germany}
\affiliation{Group of Applied Quantum Technologies (AQuT.), Friedrich-Alexander-Universität Erlangen-Nürnberg (FAU), Germany}
\author{Birgit Kallinger}
\affiliation{Fraunhofer Institute for Integrated Systems and Device Technology  (IISB), Germany}
\author{Florian Kaiser}
\affiliation{3rd Institute of Physics and Stuttgart Research Center of Photonic Engineering (SCoPE), University of Stuttgart,70569 Stuttgart, Germany}
\affiliation{Center for Integrated Quantum Science and Technology (IQST), Germany}
\author{Patrick Berwian}
\affiliation{Fraunhofer Institute for Integrated Systems and Device Technology  (IISB), Germany}
\author{Durga B. R. Dasari}
\affiliation{3rd Institute of Physics and Stuttgart Research Center of Photonic Engineering (SCoPE), University of Stuttgart,70569 Stuttgart, Germany}
\affiliation{Center for Integrated Quantum Science and Technology (IQST), Germany}
\author{Jochen Friedrich}
\affiliation{Fraunhofer Institute for Integrated Systems and Device Technology  (IISB), Germany}
\author{Roland Nagy}
\email{roland.nagy@fau.de}
\affiliation{Group of Applied Quantum Technologies (AQuT.), Friedrich-Alexander-Universität Erlangen-Nürnberg (FAU), Germany}

\date{\today}

\begin{abstract}
A distributed quantum network would require quantum nodes capable of performing arbitrary quantum information protocols with high fidelity. So far the challenge has been in realizing such  quantum nodes with features for scalable quantum computing. We show here that using the solid-state spins in 4H-Silicon Carbide (4H-SiC) such a goal could be realized, wherein a controlled generation of highly coherent qubit registers using nuclear spins is possible. Using a controlled isotope concentration and coherent control we perform here atomistic modeling of the central spin system formed by the electron spin  of a silicon vacancy color center ($V_{Si}^-$-center) and the non-interacting nuclear spins. From this we lay out conditions for realizing a scalable nuclear-spin ($^{13}C$ or $^{29}Si$) register, wherein independent control of the qubits alongside their mutual controlled operations using the central electron spin associated to the  $V_{Si}^-$-center in 4H-SiC are achieved. Further, the decoherence and entanglement analysis provided here could be used to evaluate the quantum volume of these nodes. Our results mark a clear route towards realizing scalable quantum memory nodes for applications in distributed quantum computing networks and further for quantum information protocols.
\end{abstract}

\maketitle


\section{\label{sec:level1}Introduction}

\begin{figure*}[t]
    \centering    \includegraphics[width=0.95\textwidth]{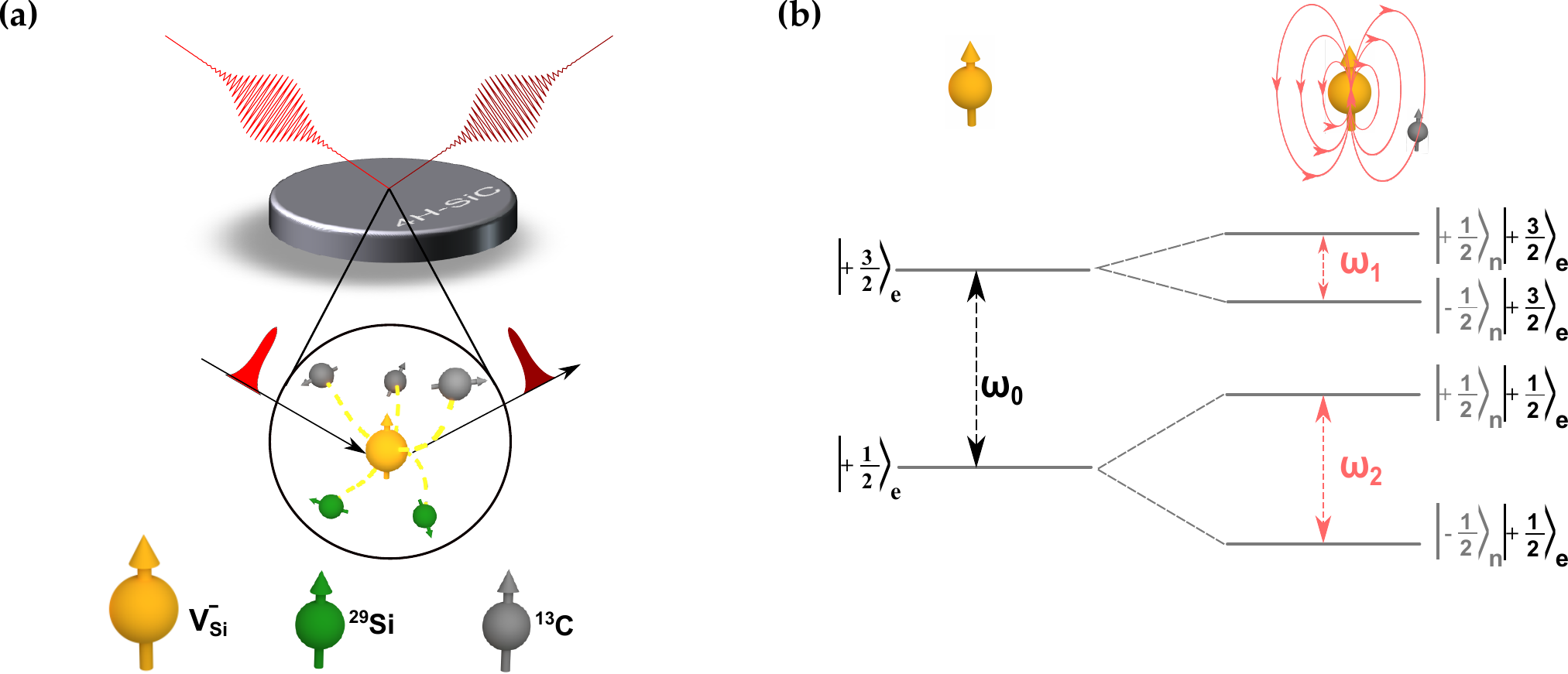}
    \caption{(a) Schematic representation of a Quantum Memory Node interacting with incoming and outgoing photons. The QMN is made of an electron-nuclear spin system as shown in the illustration. (b) Energy level scheme for a two spin subspace ($\ket{3/2}$ and $\ket{1/2}$) of $V_{Si}^-$-center without and with hyperfine coupling to a nuclear spin in the vicinity.}
    \label{fig1Part1}
\end{figure*}

The realization of a distributed quantum computing network, in which local Quantum Memory Nodes (QMN) are connected over long distances via optical photons, is an outstanding challenge in the field of applied quantum technologies \cite{ref29,ref31}. The key building block for the realization of a quantum computing network is a QMN representing a hybrid quantum system composed of qubits with different functionalities. They include qubits that form an interface with the flying qubits (photons) and other kind that allows for the processing of quantum information carried by the photons [Fig.  \ref{fig1Part1}(a)]\cite{ref22}. Over the last decade, atomistic spin-defects in solids have emerged as a potential candidate in realizing such QMN \cite{ref1,ref2,ref3,ref4}. Many components for the realization of a QMN have already been demonstrated in great details for NV and SiV color center in diamond. These realizations include long lived quantum memory qubits realized through the lattice nuclear spins \cite{ref3,ref5,ref6}, and spectrally stable optical transitions of the defect electron spin \cite{ref7,ref8,ref9}. Furthermore, demonstrations of embedding them into nano-photonic structures \cite{ref10,ref11} and a small-scale quantum network \cite{ref12} has also been achieved. Despite this progress QMN’s with scalable memory registers and at the same time displaying high cooperativities in photonic structures has not been shown so far. We address this challenge and report how the Silicon vacancy color centers ($V_{Si}^-$-centers) in 4H-SiC, with controlled doping allows for the realization of a scalable QMN {\cite{ref13,ref14,ref15,ref16}}.

The realization of a QMN requires an optically active quantum system and coherently controllable long lived memory qubits. Therefore, our analysis considers a single $V_{Si}^-$-center as a control and communication qubit surrounded by $^{13}C$ and $^{29}Si$ nuclear spins as memory qubits {\cite{ref17,ref18,ref19,ref20,ref21}}.  Electron-nuclear spin interaction is modeled by a central spin model wherein, a single $V_{Si}^-$-center couples to surrounding $^{13}C$ and $^{29}Si$ nuclear spin qubits through hyperfine interaction, which is described by the following Hamiltonian.

\begin{equation}
H=D S_z^2+ \omega_0 S_z + \omega_n I+\vec{S}{\bf A}{\vec I}.
\label{eqn1}
\end{equation}
The first term in Eq. (\ref{eqn1}) describes the ground state Zero Field Splitting (ZFS), the second and third term describe the electron and nuclear Zeeman interaction with an external magnetic field respectively. Depending on the type of the nuclear spin, the Zeeman interaction differs significantly as the gyro magnetic ratios for the carbon and silicon nuclear isotopes are  $\gamma_n=10.71$ $MHz/T$ ($^{13}C$), and $\gamma_n=-8.46$ $MHz/T$ ($^{29}Si$) respectively \cite{ref26}. The last term describes the hyperfine interaction, where ${\bf A}$ is the hyperfine tensor with elements $A_{ij}=(\mu_0 \hbar^2\gamma_e \gamma_n) (3r_ir_j-\delta_{ij})/(4\pi r^3 ) $ \cite{ref25}. With the magnetic field aligned along the quantization axis of the defect center we can safely neglect the $V_{Si}^-$-center electron spin flip terms i.e., the $S_x$ and $S_y$ terms in this hyperfine interaction. With this simplification the hyperfine tensor ${\bf A}$ becomes a vector with components $\vec{A} = \lbrace A_{zx}, A_{zy}, A_{zz} \rbrace$. Incorporating this into the nuclear spin Hamiltonian and by projecting it onto the $S^z$ basis of the electron spin, it reduces to 

\begin{equation}
    H^n= \ketbra{\pm 3/2}{\pm 3/2}    \otimes H^n_{\pm 3/2}+ \ketbra{\pm 1/2}{\pm 1/2} \otimes H^n_{\pm 1/2}.
\label{eqn2}
\end{equation}
As noted above the $V_{Si}^-$-center electron spin is a four-level system (FLS) with eigen basis states $\ket{\pm 3/2}_e$ and $\ket{\pm 1/2}_e$. The nuclear dynamics conditioned on the electron spin state is determined by 
$H^n_{\pm 3/2} =\omega_n I_z \pm 3/2 (A_\parallel I_Z + A_\perp I_x) $, and 
$H^n_{\pm 1/2} =\omega_n I_z \pm 1/2 (A_\parallel I_Z + A_\perp I_x) $. Instead of working in the full FLS, we will restrict the dynamics to the effective $\ket{3/2}_e$ and $\ket{1/2}_e$ two-level subspace. This is realized by an applied static magnetic field with 500 G aligned along the quantization axis of the $V_{Si}^-$-center. The frequency of applied microwave addresses the nuclear spin transitions corresponding to the $V_{Si}^-$-center electron spin state of $\ket{3/2}_e$ and $\ket{1/2}_e$.
\begin{figure*}[t]
    \centering
    \includegraphics[width=\textwidth]{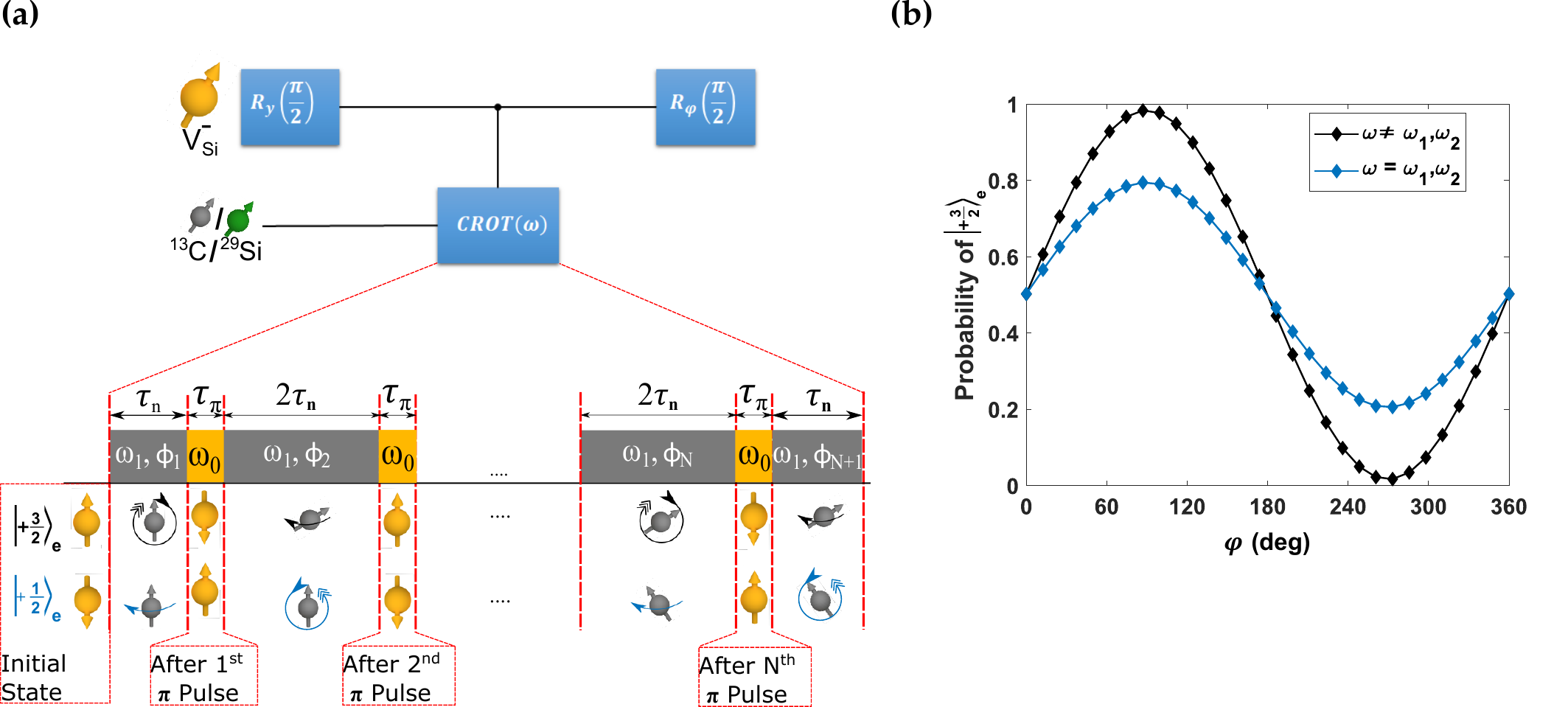}
    \caption{ (a) $DDrf$ pulse sequence used for detecting and controlling nuclear spins with interleaved  dynamic decoupling on the $V_{Si}^-$-center to preserve the electron spin coherence. The sequence also depicts the nuclear spin being driven for a {2$\tau_n$ period between electron spin $\pi$ pulses}. The dynamic decoupling sequence uses $N$  $\pi$ pulses on the  $V_{Si}^-$-center electron spin.{ The phase of the $n^{th}$ $rf$ signal which drives the nuclear spin varies as $\phi_n$.(An explanation regarding $\phi_n$ is shown in main text and in Methods section)}   (b) The result of {the final electron spin state population is readout with varying phase ($\varphi$) of our final $\pi/2$ pulse as shown in (a). In the scenario when the $rf$ frequency $\omega$  is detuned from the nuclear spin levels ($\omega_1 $ or $\omega_2$), an inversion of spin state population from $\ket{3/2}_e \Leftrightarrow  \ket{1/2}_e$  is observed. Whereas when the $rf$ frequency $\omega$ is in resonance with  $\omega_1$ or $\omega_2$ (as shown in (a) for $\omega =\omega_1$) a reduction of the  probability amplitude between $\ket{3/2}_e \Leftrightarrow  \ket{1/2}_e$  is observed.}}
    \label{fig1Part2}
\end{figure*}

The realization of a QMN requires coherent control of single nuclear spins via electron-nuclear gate sequences. The main challenge in implementing these sequences is to maintain spin coherence on the electron-spin and avoid unwanted crosstalk between nuclear-spin qubits. To address this problem, a control sequence is required which preserves the spin coherence of the $V_{Si}^-$-center and at the same time also performs a selective rotation on the nuclear-spin qubits. This could be achieved by a  pulse sequences which involves a two-qubit gate based upon phase-controlled radio-frequency ($rf$) which drives the nuclear spins interleaved with dynamical decoupling (DD) of the electron spin \cite{ref6}. $DDrf$ sequences enable the control of multiple nuclear spin qubits while maintaining the spin coherence on the electron spin ($V_{Si}^-$-center) and reducing their crosstalk. The concept of $DDrf$ sequences is based on selective two-qubit gates. Therefore, hyperfine interaction is utilized which couples each nuclear spin to the $V_{Si}^-$-center. This hyperfine interaction depends on the position of the nuclear spin qubit relative to the $V_{Si}^-$-center \cite{ref24,ref25}. 

We further control the nuclear spins through $rf$ control wherein the applied frequency $\omega$, phase $\phi$, and the amplitude (Rabi frequency) $\Omega$ are adjusted to allow for independent control and coupling of a given nuclear spin to the $V_{Si}^-$-center electron spin. For example to address a specific nuclear spin when the $V_{Si}^-$-center is in the spin state $\ket{3/2}_e$ , we set $\omega=\omega_1=\sqrt{(\omega_L-3A_\parallel/2)^2+9 A_\perp^2/4 }$ or at $\omega=\omega_2=\sqrt{(\omega_L-A_\parallel/2)^2+ A_\perp ^2/4 }$, when the $V_{Si}^-$-center is in the state $\ket{1/2}_e$. Further, for a very low (negligible) perpendicular coupling $A_\perp$ and a Rabi frequency $\Omega \ll (\omega_2-\omega_1)$, the  conditional nuclear Hamiltonian gets further simplified such that the total Hamiltonian given in Eq. (\ref{eqn3}) takes the form 

\begin{equation}
\begin{aligned}
H^n= \ketbra{\pm 3/2}{\pm 3/2}    \otimes \Omega(cos(\phi)I_x +sin(\phi)I_y)+\\
 \ketbra{\pm 1/2}{\pm 1/2} \otimes (\omega_2-\omega_1)I_Z.
\label{eqn3}
\end{aligned}
\end{equation}

One can see from the equation above that there is a  rotation of the nuclear spin  {about x-y plane when $V_{Si}^-$-center electron spin is conditioned to be in the state $\ket{3/2}_e$} and a phase evolution {of nuclear spin about z-axis} when {$V_{Si}^-$-center electron spin is conditioned to be in the state $\ket{1/2}_e$}. {The conditional phase evolution is governed by the detuning ($\omega_2-\omega_1$) of the applied $rf$ pulse ($\omega=\omega_1$) with the nuclear spin energy difference ($\omega_2$) when the electron spin is at $\ket{1/2}_e$ (as depicted in Fig. \ref{fig1Part1}(b)), as shown in Eq. (\ref{eqn3})}. A pulse sequence which utilizes Eq. (\ref{eqn3}) to identify and coherently control single nuclear spin qubits via conditional rotations is shown in [Fig. \ref{fig1Part2}(a)]. The sequence begins by preparing the $V_{Si}^-$-center electron spin in a superposition state, $1/\sqrt{2}[\ket{1/2}_e+\ket{3/2}_e]$, using a $\pi_y/2$ pulse. After this a controlled rotation $CROT(\omega)$ sequence is applied [insert Fig. \ref{fig1Part2}(a)]. When the $rf$ frequency $\omega$ is resonant with a given nuclear spin qubit ({$\omega = \omega_1$ or $\omega_2$}), the $CROT(\omega)$ acts as an electron-nuclear $DDrf$ two-qubit conditional phase rotation as shown in Eq. (\ref{eqn3}). { The phase of each} $rf$ pulse on the nuclear spin is labelled as {$\phi_{n=1,\cdots N+1}$ with $N$ being the total number of $\pi$ pulses on $V_{Si}^-$-center electron spin. Whereas, $N+1$ denotes the  total number of $rf$ pulses on the nuclear spin.} {If the initial state of the electron spin is $\ket{3/2}_e$ and the $rf$ frequency $\omega = \omega_1$, the nuclear spin will undergo a rotation about x-y plane. The nuclear spin evolution is followed by a $\pi$ pulse which flips the electron spin state from $\ket{3/2}_e$ to $\ket{1/2}_e$. Subsequently the nuclear spin will undergo a free evolution about the z-axis. This sequence will be repeated until the $N+1$ $rf$ pulse on the nuclear spin. }

{
\begin{equation}
\begin{split}
\phi_{1} = \phi_{initial}\\
\phi_{2} = \phi_{1} + \phi_{\tau_n} + \pi\\
\phi_{l+2} = \phi_{l} + 2\phi_{\tau_n}\\
\phi_{N+1} = \phi_{N-1} + \phi_{\tau_n}\\
\label{eqn4New}
\end{split}
\end{equation}
}

{The phase of each applied nuclear spin $rf$ pulse ($\phi_{n=1,\cdots ,l,l+1,l+2,\cdots,N+1}$) is given by Eq. (\ref{eqn4New}). The phase of the first nuclear spin $rf$ pulse can be freely set to $\phi_{initial}$. Correspondingly, the phase of consecutive $rf$ nuclear spin pulses is calculated depending on the phase obtained due to the free evolution $\phi_{\tau_n} = (\omega_2-\omega_1)\tau_n$ for a time period of $\tau_n$ as shown in  Eq. (\ref{eqn4New}). Furthermore a $180^\circ$ shift is required between consecutive nuclear spin $rf$ pulses to alter their rotation direction depending on the electron spin state. A detailed explanation on nuclear spin phase evolution is described in the Methods section.} 

 The final readout of the sequence is done by $\pi_\varphi/2$ –pulse on the electron spin {with varying phase $\varphi$}. The corresponding results are shown in Fig. \ref{fig1Part2}(b). {The reduction in the probability amplitude of the final readout is governed by loss of coherence  due to the scenario where $\omega = \omega_1$ or $\omega_2$. This effect is used for the detection of nuclear spins in the vicinity of a $V_{Si}^-$-center.} Using the sequence shown in Fig. \ref{fig1Part2}(a), we perform the numerical simulations to detect the nuclear spin qubits. For negligible $A_\perp$ ($A_{xz}$ or $A_{yz}$) coupling, the fidelity of detecting the nuclear spin can be maximized for the number of $\pi$ pulses $N \gg \pi/(2\Omega\tau_n)$. By varying the driving frequency $\omega$  the sequence is capable of detecting randomly positioned $^{13}C$ and $^{29}Si$ nuclear spin qubits. We would like to note that all nuclear spin qubits show two hyperfine transitions positioned at {$\omega = \omega_1$ and $\omega = \omega_2$} {(due to the driving of nuclear spins at the corresponding spin state of the $V_{Si}^-$-center) }. This leads to an additional control which we discuss later.
\begin{figure}[t]
    \centering
    \includegraphics[width=0.5\textwidth]{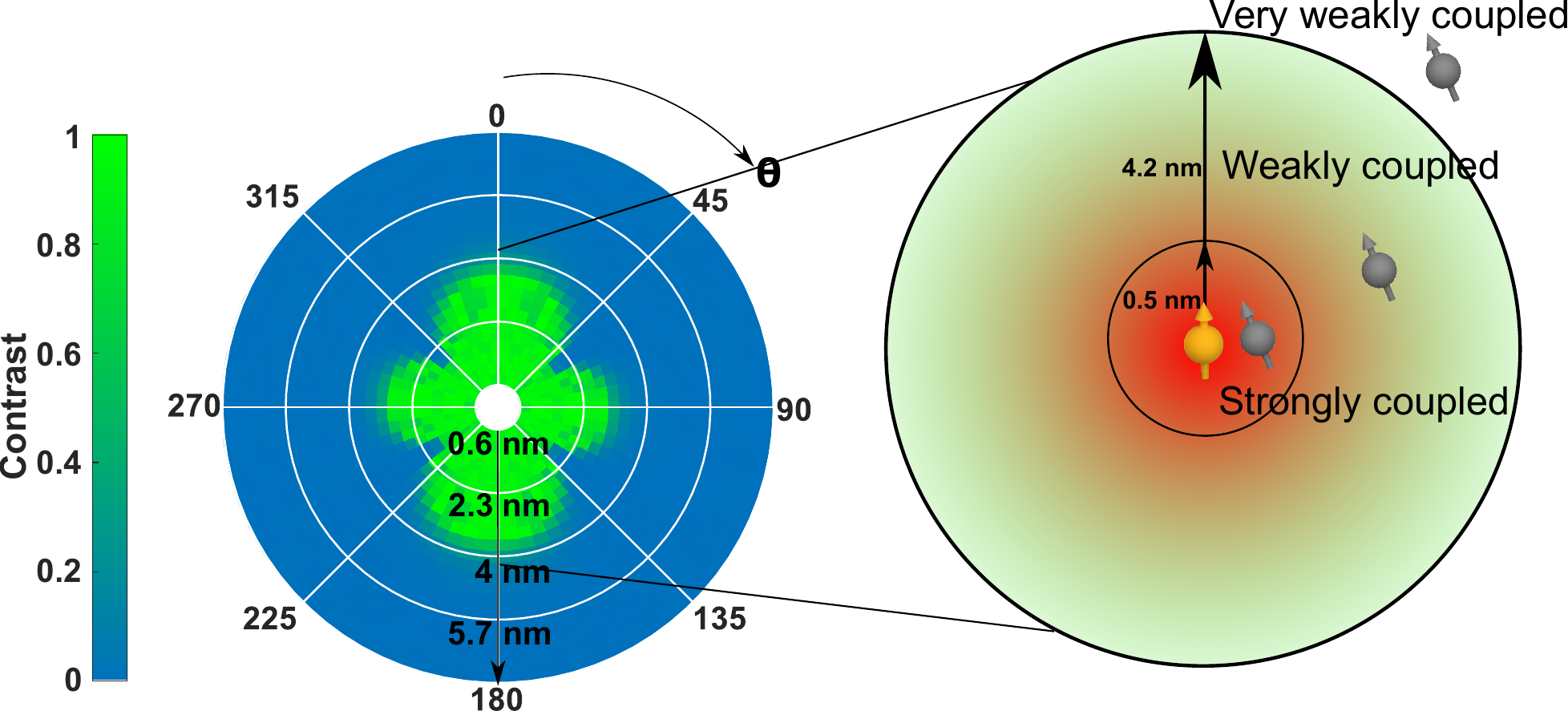}
    \caption{The contrast in controlling a nuclear spin in the vicinity of a $V_{Si}^-$-center electron spin is shown as a function of the relative position of nuclear spins in a polar plot. The nuclear spins that are closer to the $V_{Si}^-$-center electron spin are coupled strongly with a coupling strength $A>\frac{1}{T_2^*}$. The nuclear spins that are further away from the central $V_{Si}^-$-center electron spin require more driving time in order to access them owing to a very low coupling strength. The accessibility of such very weakly coupled nuclear spin are limited by the $V_{Si}^-$-center electron spin coherence.}
    \label{fig2}
\end{figure}

\begin{figure}[t]
    \centering
    \includegraphics[width=0.5\textwidth]{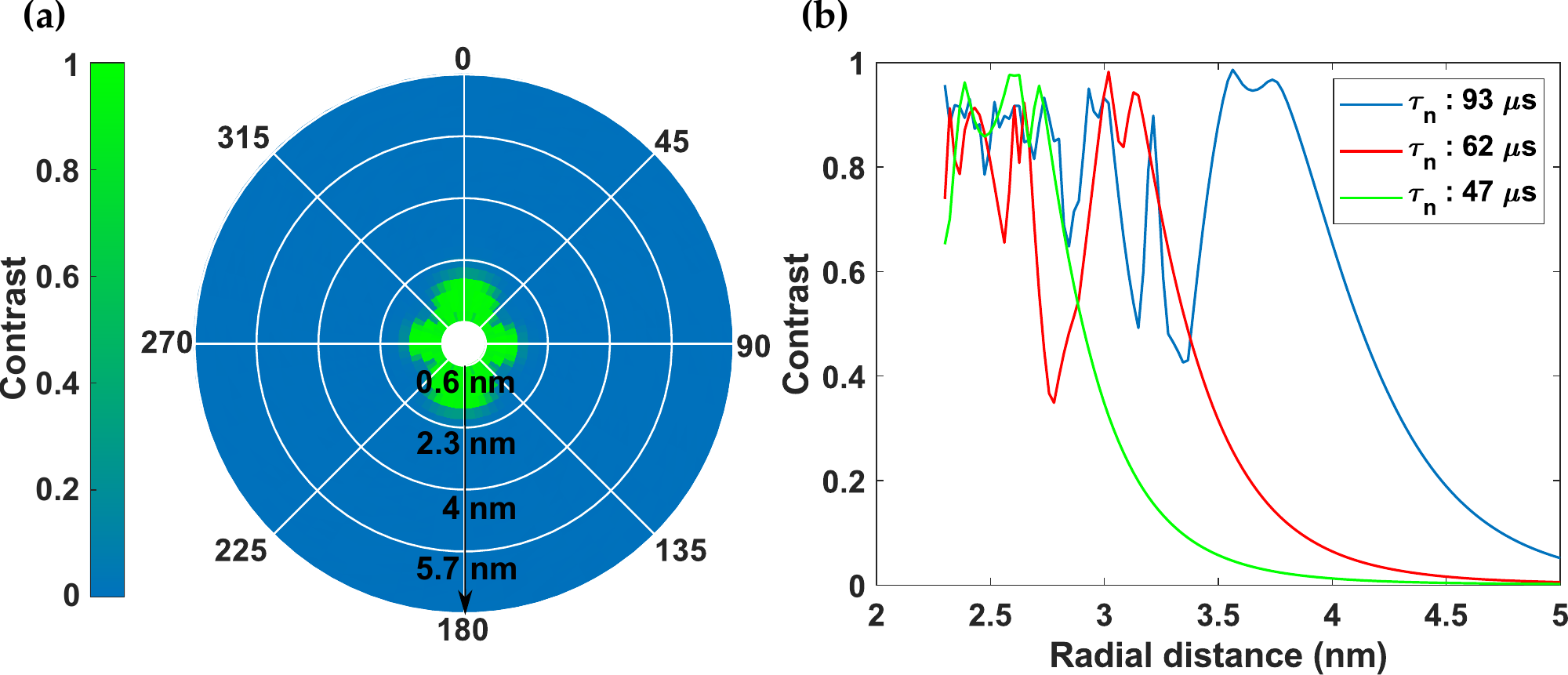}
    \caption{(a)  The polar plot of the nuclear spin contrast for a fixed values of  $N=20$ and $\tau_n=46$ $\mu s$ is depicted. Nuclear spins closer to the $V_{Si}^-$-center electron spin ($\sim 2$ nm) are accessible unlike those shown Fig. \ref{fig2}. (b) The radial dependence on accessing a nuclear spin depending on the contrast for $\theta = 0$ depicts that the radial distance within which the nuclear spins are accessible reduces with decreasing $\tau_n$.}
    \label{fig3}
\end{figure}

Upon turning onto the atomistic modeling of the nuclear spin bath from the physical parameters associated to the SiC lattice, a volume of $680$ $nm^3$ is considered with the position of the $V_{Si}^-$-center  set at the origin for the lattice space. This corresponds to approximately $8200$ unit cells of the lattice. Further, we consider a natural isotopic abundance of $^{13}C$ and $^{29}Si$ nuclear spins to be $1.1\%$ and $4.7\%$ respectively. With this isotopic concentration, the average distance between any two nuclear spins is larger than $6$ nm, indicating a weak (negligible) intra-nuclear spin interactions. For this reason in the reminder of the discussion we consider all the nuclear spins of the bath are non-interacting and any interaction between them will be mediated by the $V_{Si}^-$-center electron spin alone. Further, the average distance at natural isotopic abundance of the $V_{Si}^-$-center electron spin to the nearest nuclear spins is in the order of  $0.5$ nm, leading to the coupling strength in the range of  $Hz$ to $\sim 200$ $kHz$. To optimize the maximum accessible number of controllable nuclear spin qubits as potential quantum memories we perform a parametric analysis by employing the sequence from Fig. \ref{fig1Part2}(a). An important property, which influences the outcome of the measurement protocol, is the concentration of nuclear spins that dictates the total number of isotopes in the vicinity of a $V_{Si}^-$-center  \cite{ref27}. To analyze the influence of the isotopic concentration on the maximum number of controllable nuclear spins we initially simulate the measured contrast of a single $^{13}C$ and $^{29}Si$ nuclear isotope at all possible positions around a $V_{Si}^-$-center.{ The contrast in amplitude as shown in Fig. \ref{fig1Part2}(b) is analyzed using the pulse sequence from  Fig. \ref{fig1Part2}(a) for each positions (radial $r$ and angular $\theta$ coordinates) of a nuclear spin from $V_{Si}^-$-center}. The analysis was initially performed for a long nuclear spin driving time of  $\tau_n=93$ $\mu s$ with $N=100$ which amounts to the total driving time of $2N\tau_n = 18.6$ $ms$ [Fig. \ref{fig2}]. However, the choice of nuclear spin driving time is restricted by the $V_{Si}^-$-center electron spin coherence. Therefore, this analysis was also performed for a short nuclear spin driving time $\tau_n=46$ with $\mu s$ $N=20$ (total driving time of $2N\tau_n = 1.8$  $ms$)[Fig. \ref{fig3}(a)]. However, the simulation shows that the sensing volume within which a nuclear spin can be detected is reduced. This behaviour arises due to the fact that the maximum radial distance until which nuclear spins are accessible (for $\theta=0^{\circ}$ or $\theta=180^{\circ}$) reduces with decreasing nuclear spin driving time $\tau_n$ as depicted in Fig. \ref{fig3}(b). The rise and drops of the contrast in Fig. \ref{fig3}(b) is suspected to be the cause of the revival of the electron spin coherence when the nuclear spin exhibits a complete rotation. The contrast drop beyond a certain radial distance in Fig. \ref{fig3}(b) indicates that a longer driving time is necessary to access these nuclear spin. 
To analyze the usability of nuclear spins we set two criteria. First (i) we consider only nuclear spin qubits at a lattice position where the contrast is larger than $> 0.5$. Second (ii) we consider only nuclear spin qubits which can be independently controlled (independent $rf$ driving frequency). Exploiting the  interdependence of contrast and $rf$ driving frequency $\omega$ we can statistically analyze the controllable nuclear spin qubits by looking for nuclear spins that satisfy the aforementioned criteria. If we assume a natural abundance of $1.1\%$ of $^{13}C$ and $4.7\%$ of $^{29}Si$, the total number of accessible nuclear spins which meet the first criteria are $N_{Qubits} = 161$ for $N=100$ with $\tau_n=93$ $\mu s$ and $N_{Qubits} = 23$ for $N=20$ with $\tau_n=46$ $\mu s$  respectively. Unfortunately, a high amount of accessible nuclear spins leads to driving of multiple nuclear spins within an identical driving frequency and henceforth to a lower amount of controllable nuclear spin qubits that could exhibit independent driving [see Methods]. Hence upon following both the criterias at natural abundance of isotopes the $N_{Qubits} = 16$ for $N=100$ with $\tau_n=93$ $\mu s$ and $N_{Qubits} = 14$ for $N=20$ with $\tau_n=46$ $\mu s$ respectively. 
\begin{figure*}[t]
    \centering
    \includegraphics[width=0.8\textwidth]{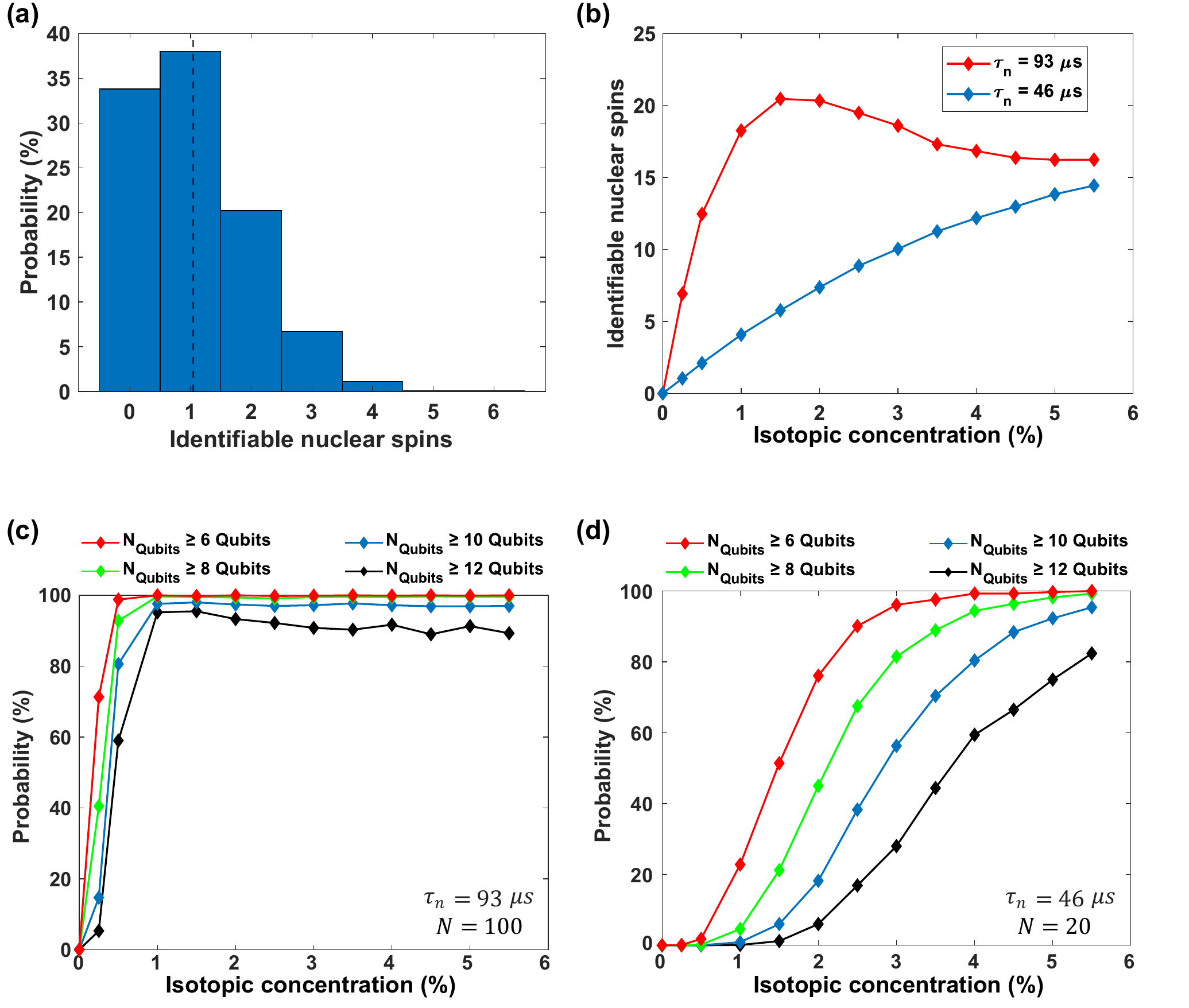}
    \caption{(a) A histogram on the number of nuclear spins that are independently accessible is plotted for a $[^{13}C] = 0.125\%$ and $[^{29}Si] = 0.125\%$ along with the mean accessible nuclear spins (dotted line). The analysis is carried out for 1000 different distribution of nuclear spin around the central $V_{Si}^-$-center electron spin. (b) The mean number of accessible nuclear spins for varying isotopic concentrations are depicted for  two configurations. The plot reveals a possibility to access 20 nuclear spins at total isotopic concentration of $1.5\%$ for  $N=100$ with $\tau_n=93$ $\mu s$. (c) and (d) represents the probability in accessing more as 6 qubits to 12 qubits with varying isotopic concentration for  $N=100$ with $\tau_n=93$ $\mu s$ and  $N=20$ with $\tau_n=46$ $\mu s$ respectively. }
    \label{fig4}
\end{figure*}
In our analysis, we are varying the concentration of $^{13}C$ and $^{29}Si$ between $0.25\%$ and $5.8\%$. The variation of isotopic concentration is such that the proportion of $^{13}C$ is same as $^{29}Si$ until $^{13}C$ concentration reaches its natural abundant concentration ($1.1\%$), beyond which the concentration of $^{29}Si$ is varied. We are performing a statistical analysis over $1000$ distributions for all given concentration of $^{13}C$ and $^{29}Si$. All nuclear spin qubits which do not fulfill both aforementioned criterias are excluded in our analysis. 

\begin{figure}[t]
    \centering
    \includegraphics[width=0.5\textwidth]{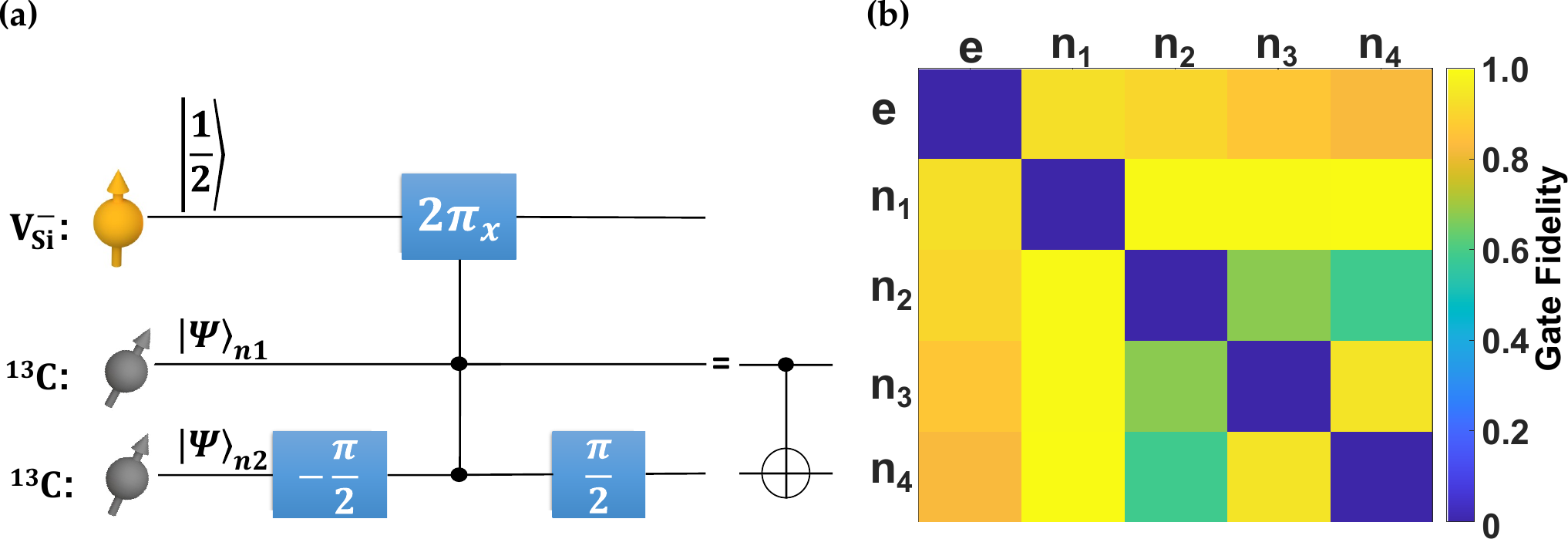}
    \caption{(a) The $V_{Si}^-$-center electron spin controlled $CNOT$ gate between two uncoupled nuclear spins. The controlled rotation of $V_{Si}^-$-center electron spin is performed for $\ket{1/2}_{n1}\ket{1/2}_{n2}$ of nuclear spins  (b) For a random choice of nuclear spins at a $^{13}C$ concentration of $1\%$, we show the entangling gate fidelities among various nuclear spins. In the above simulation the couplings of the nuclear spins to the central spin are: $\lbrace A_{zz} = 12.5\,kHz,\,-3.8\,kHz,\,-18.9\,kHz,\,13.4\,kHz$ and $A_{xz} =  2.3\,kHz,\,5.1\,kHz,\,-13\,kHz,\,9\,kHz\rbrace$ respectively.}
    \label{fig5}
\end{figure}

The total number of accessible nuclear spins at an isotopic concentration of $0.25\%$ (with the sum of [$^{13}C$] = $0.125\%$ and [$^{29}Si$] = $0.125\%$) considering the two aforementioned criterias is shown as a histogram in Fig. \ref{fig4}(a). Wherein the mean accessible nuclear spins and the probability distributions for our analysis are derived. The mean accessible nuclear spins for a sum concentration of $^{13}C$ and $^{29}Si$ isotopes are depicted in Fig. \ref{fig4}(b). The curve saturates for $\tau_n = 93$ $\mu s$ and $N=100$ at an isotopic concentration of $1.5\%$. At the mentioned isotopic concentration there are 49 nuclear spin qubits within the sensing volume which meets criterion (i) for $\tau_n = 93$ $\mu s$ and $N=100$. Meanwhile, for $\tau_n = 46$ $\mu s$ and $N=20$  there are only 8 nuclear spin qubits within the sensing volume that meets criterion (i). Owing to more nuclear spins within the sensing volume for $\tau_n = 93$ $\mu s$ and $N=100$ the effect of non independent driving within the sensing volume starts to dominate as shown at $1.5\%$ in Fig. \ref{fig4}(b). Hence the mean identifiable nuclear spins for $\tau_n = 93$ $\mu s$ with $N=100$ reduces until the point where the effect of increasing nuclear spins within the sensing volume and deteriorating controllable nuclear spins due to non independent driving cancels out. The curve hence saturates at naturally abundant concentration of isotopes [see Methods]. This results shows the importance of tailored isotopically produced 4H-SiC epitaxial layers to maximize the nuclear spin qubit access. A similar behaviour is also noticeable for $\tau_n = 46$ $\mu s$ with $N=20$, but in this case at an much higher isotopic concentration owing to a smaller sensing volume.


Based on the analysis from Fig. \ref{fig4}(a), it is also clearly indicated in Figs. \ref{fig4}(c) and \ref{fig4}(d) that $N_{Qubits} > 10$ are achievable for both scenarios where $\tau_n=93$ $\mu s$ and $\tau_n=46$ $\mu s$ respectively. The probability distribution can hence be adjusted for one given isotopic concentration by adjusting the nuclear spin driving time. However, the  choice of $\tau_n$ and $N$ also depends on the spin coherence property corresponding to the given isotopic concentration. For further understanding of the relation between spin coherence and isotopic concentration, an investigation is conducted through the method of Cluster Correlation Expansion (CCE) \cite{ref26, ref28} [see Methods]. The investigation is realized with static magnetic field of 500 G. The analysis reveals a possibility of preserving the electron spin coherence up to $~2$ $ms$ at an isotopic concentration of $1.0\%$. However the electron spin coherence deteriorates with increasing isotope concentration. Hence the choice of an optimal $rf$ driving time $\tau_n$ for an isotopically pure sample acts as an important experimental parameter that maximizes the accessible and controllable nuclear spin qubits per quantum memory node. For instance, a $DDrf$ sequence with $\tau_n=93$ $\mu s$ and  $N=100$ demands the electron spin coherence time  $>18.6$ $ms$  while the configuration of $\tau_n=46$ $\mu s$ and  $N=20$ demands the electron spin coherence time  $>1.86$ $ms$.

To determine the efficiency of a quantum computer, new indicators such as quantum volume were introduced recently \cite{ibm}. These measures determine the actual number of high fidelity quantum bits that are useful for computing among the many physical qubits available (e.g., $20$ on average in our case). To evaluate this measure one needs to perform $SU(4)$ operations on any two qubits in the quantum register, and this needs the ability to perform a $CNOT$ gate between any two qubits among the detected spins. 
For this we have performed the analysis on the generation of a $CNOT$ gate between pairs for nuclear spin qubits which potentially plays a pivotal role in generation of a maximally entangled state \cite{ref6}. The $CNOT$ gate between a $V_{Si}^-$-center and nuclear spins ($n_1, n_2, n_3$ and $n_4$) can be achieved by $DDrf$ sequence through a controlled rotation of a nuclear spin by $\pi/2$ [Fig \ref{fig1Part2}(a)] \cite{ref6}. The $CNOT$ gate between a pair of nuclear spin qubits can only be achieved through the intervention of an electron spin qubit. The sequence to generate a CNOT gate between a pair of nuclear spin qubits is shown in Fig. \ref{fig5}(a) \cite{ref32}. The $\pi/2$ gate operation on nuclear spins are performed through the $DDrf$ sequence as shown in Fig. \ref{fig1Part2}(a). The analysis is conducted at an isotopic spin bath concentration of $1\%$ with a random spatial nuclear spin bath configuration. The result of the simulation as shown in Fig. \ref{fig5}(b) reveals a  fidelity $>90\%$ in the generation of a $CNOT$ gate between two nuclear spin qubits mediated by a $V_{Si}^-$-center electron spin. 

In conclusion, our analysis showed a notable chance of accessing $20$ nuclear spin qubits at an isotopic $^{13}C$ and $^{29}Si$ concentrations in the range of $1 - 1.5\%$ [Fig. \ref{fig4}(b)]. Hence a control on isotopic purity is necessary in order to maximize the number of nuclear spin qubits. A minimum of 6 memory qubits can be detected with significant probability of $>20\%$ at an isotopic concentration of $1.0\%$. A way to increase the register size further is to adjust the experimental driving time $\tau_n$. As shown in Fig. \ref{fig2}(a) longer driving time $\tau_n$ increases the sensing volume thereby allowing weakly interacting nuclear spin qubits to be identified. However, the choice of $\tau_n$ is also limited by the electron spin coherence property which further limits the sensing volume. Hence, for a low isotopic concentration it would be beneficial to use a long nuclear spin driving time $\tau_n$ and a high number of repetitions $N$ to increase the sensing volume. Unfortunately, electron spin coherence sets an upper limit for the nuclear spin driving time and number of repetitions. For a higher isotopic concentration like natural abundance, it would be beneficial to drive the nuclear spins with a shorter nuclear spin driving time in order to reduce  non independent driving. Therefore, our presented work gives a detailed answer to the question how many memory qubits can be identified and controlled via a single $V_{Si}^-$-center. Further with the ability to achieve high-fidelity entangling gates among the nuclear spins and the decoherence analysis performed one should be able to estimate the Quantum volume as a function of the isotopic concentration in these materials.

\begin{acknowledgments}
We acknowledge financial support by the Federal Ministry of Education and Research (BMBF) project QMNDQCNet, and by the Fraunhofer Start Project “Quantum Computing”. DDBRao would like to acknowledge the support by DFG (FOR2724).
\end{acknowledgments}

\newpage

\bibliographystyle{ieeetr}
\bibliography{NV-Centers.bib, Software.bib, Batteries.bib, Books.bib}
\clearpage
\newpage
\widetext

\begin{center}
    \textbf{\large Scalable quantum memory nodes using nuclear spins in Silicon Carbide - Supplementary Information}
\end{center}

\setcounter{equation}{0}
\setcounter{figure}{0}
\setcounter{table}{0}
\setcounter{page}{1}
\setcounter{section}{0}
\makeatletter
\renewcommand{\theequation}{S\arabic{equation}}
\renewcommand{\thefigure}{S\arabic{figure}}

\section{Lattice structure of 4H-SiC}\label{App:A1}
The material platform silicon carbide has variety of polytypes [S1]. However the 4H polytype of Silicon Carbide is of interest to our application. The 4H polytype of SiC has an hexoganality of $50\%$. The unit cell consists of an effective number of 4 Si and 4 C atoms. The parameters of unit cell are
\begin{equation}
\begin{aligned}
a = b = 0.3073\,nm\\
c = 1.0053\,nm\\
\alpha = \beta = 90 ^{\circ}\\
\gamma = 120 ^{\circ}
\label{eqn5}
\end{aligned}    
\end{equation}
Hence the lattice vectors that dictate the position of atoms are
\begin{equation}
\begin{split}
\vec{r} = n_1\vec{a_1} + n_2\vec{a_2} + n_3\vec{a_3}\\
\vec{a_1} = a\vec{e_x}\\
\vec{a_2} = a(-\cos{(\gamma)}\vec{e_x}+\sin{(\gamma)}\vec{e_y})\\
\vec{a_3} = c\vec{e_z}
\label{eqn5}
\end{split}
\end{equation}
\setcounter{figure}{6}  

\begin{figure}[t]
    \centering
    \includegraphics[width=\textwidth]{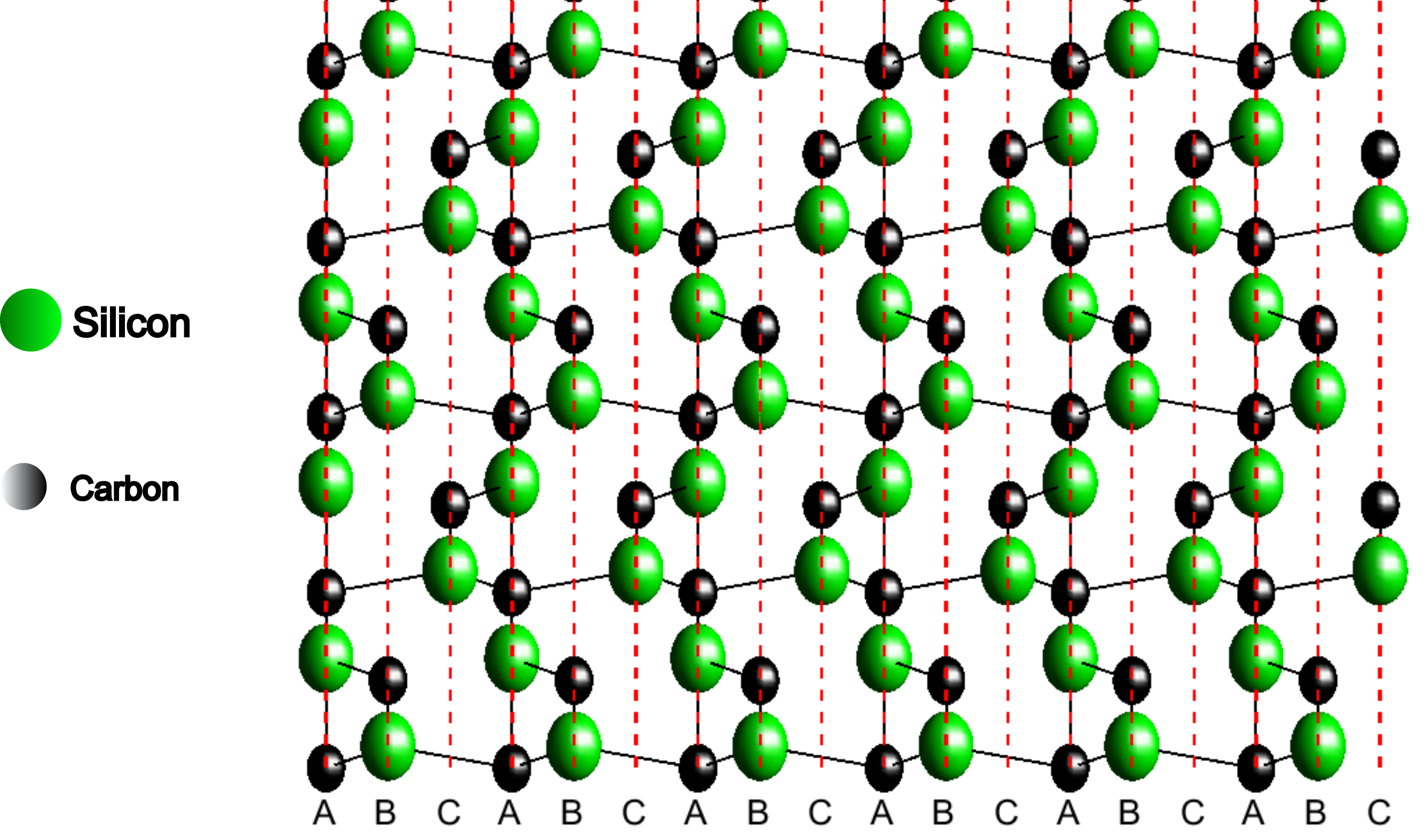}
    \caption{The lattice structure of 4H-SiC is depicted. The lattice represents hexagonality of $50\%$. A typical unit cell of 4H-SiC consist of 4 Si and 4 C atoms [S1,S2].}
    \label{fig6}
\end{figure}

\section{Spin Hamiltonian}\label{App:A2}

\subsection{Nuclear spin Hamiltonian}
The Hamiltonian for nuclear spin decides the state evolution of nuclear spin and energy level distribution within the nuclear spin through the Schrödinger equation.The Hamiltonian of the electron - nuclear spin interaction ($l=0,\,k\neq0$) and that of nuclear-nuclear spin interaction (($l\neq0,\,k\neq0$) ) is written as Eq. (\ref{eqn18}).

\begin{equation}
\begin{split}
\Tilde{A_{lk}} = \frac{(\mu_0 \hbar^2\gamma_l \gamma_k) (3\vec{r_{lk}}\bigotimes \vec{r_{lk}} - \mathbb{I})}{(4\pi r_{lk}^3 )} \\
H_{m_s}^{0,k} = \gamma_l B_0 I_z + \sum_{i=x,y,z} m_s \Tilde{A_{0k}} I_i\\
H_{m_s}^{l,k} = H_{m_s}^{0,l} + H_{m_s}^{0,k} + \sum_{i,j=x,y,z} I_i \Tilde{A_{lk}} I_j\\
\label{eqn18}
\end{split}
\end{equation}

Upon considering a single nucleus interacting with a  $V_{Si}^-$-center, the Hamiltonian for the nuclear spin when driven by $rf$ field as shown in Eq. (\ref{eqn6}) is affected by the spin state of the $V_{Si}^-$-center [S2]. The equation in Eq. (\ref{eqn6}) represents the nuclear spin Hamiltonian in the presence of a driving microwave field which is represented as $2\Omega \cos{(\omega t + \phi)} I_x$. The first term refers to the nuclear Zeeman splitting between the nuclear spins states. The second and third term represents the magnetic dipole dipole interaction between the $V_{Si}^-$-center electron spin and nuclear spin. Without loss of generality, we assume the  $V_{Si}^-$-center electron spin and nuclear spin to be available on x-z plane. The spin flip of $V_{Si}^-$-center electron spin due to dipole dipole interaction between the color center spin and nuclear spin is neglected. Hence the magnetic dipole interaction considers only $A_{xz}$, $A_{yz}$ and $A_{zz}$ terms. {The term $\sqrt{A_{xz}^2 +A_{yz}^2}$ is referred to as $A_\perp$ while the term $A_{zz}$ is referred to as $A_{\parallel}$.}

\begin{equation}
\begin{split}
H_{n\ket{m_s}_e}(\omega,\phi,t) = \gamma_n B_0 I_z + m_s A_{zz} I_z + m_s A_{xz} I_x\\ + 2\Omega \cos{(\omega t + \phi)} I_x \\
\label{eqn6}
\end{split}
\end{equation} 

The equation Eq. (\ref{eqn6}) can be rewritten in terms of the nuclear larmor frequency ($\omega_1$ for $m_s=1/2$ and $\omega_2$ for $m_s=3/2$) where $\beta_{m_s}$ denotes the shift in quantization axis from $\vec{e_z}$ due to magnetic dipole dipole interaction [Eq. (\ref{eqn7})].

\begin{equation}
\begin{split}
H_{n\ket{m_s}_e}(\omega,\phi,t) = \omega_{1,2}(\cos{\beta_{m_s}} I_z +\sin{\beta_{m_s}} I_x)\\ + 2\Omega \cos{(\omega t + \phi)} I_x 
\label{eqn7}
\end{split}
\end{equation}

The coordinate transformation to the shifted quantization axis will be mathematically beneficial since the time independent part of the Hamiltonian will have the $I_{z,\beta_{m_s}}$ term only. The transformed $I_{z,\beta_{m_s}}$ and $I_{x,\beta_{m_s}}$ is depicted in Eq. (\ref{eqn8}).

\begin{equation}
\begin{split}
R_{x,y,z}(\theta) = e^{(-i\theta I_{x,y,z})}\\
I_{z,\beta_{m_s}} = R_y(\beta_{m_s}) I_z R_y^T(\beta_{m_s})\\
I_{x,\beta_{m_s}} = R_y(\beta_{m_s}) I_x R_y^T(\beta_{m_s})\\
\label{eqn8}
\end{split}
\end{equation}

Hence the Hamiltonian in terms of transformed coordinates is written as Eq. (\ref{eqn9}). The following representation makes it simpler for the rotating frame analysis in upcoming section.
\begin{equation}
\begin{split}
H_{n\ket{m_s}_e}(\omega,\phi,t) = \omega_{1,2}I_{z,\beta_{m_s}} +\\ 2\Omega\cos{\beta_{m_s}} \cos{(\omega t + \phi)} I_{x,\beta_{m_s}} +\\ 2\Omega\sin{\beta_{m_s}} \cos{(\omega t + \phi)} I_{z,\beta_{m_s}}
\label{eqn9}
\end{split}
\end{equation}

\subsection{Rotating Frame Approximation}
The time dependence of the Hamiltonian challenges the analysis of nuclear spin state evolution through the time dependent Schrödinger equation. Owing to this drawback, the frame of reference for the analysis is considered to be rotating about the quantization axis with the frequency of the applied MW $\omega$. Hence if the $\omega \backsim \omega_1$ or $\omega_2$ the time dependent term ca be neglected owing to the rotating wave approximation [S3]. Therefore, the Hamiltonian in rotating frame as depicted in Eq. (\ref{eqn10}) would exhibit time independence thus making the state evolution analysis simpler. 
\begin{equation}
\begin{split}
H_{n\ket{m_s}_e}^R(\omega,\phi)  = R_z(\omega t) (H_{n\ket{m_s}_e}(\omega,\phi,t) - \omega I_{z,\beta_{m_s}}) R_z^T(\omega t)\\
\label{eqn10}
\end{split}
\end{equation}
Upon application of a MW sequence for time $t$ to control a nuclear spin corresponding to $V_{Si}^-$-center electron spin of $\ket{m_s}_e$ the unitary matrix in Eq. (\ref{eqn11}) is capable of describing the state evolution of nuclear spin.
\begin{equation}
\begin{split}
U_{m_s} (t,\omega,\phi) = e^{(-iH_{n\ket{m_s}_e}^R(\omega,\phi)t)}
\label{eqn11}
\end{split}
\end{equation}

\section{Controlled rotation}\label{App:A3}
\subsection{{\bf DDrf} pulse sequence}

\begin{figure}[t]
    \centering
    \includegraphics[width=0.8\textwidth]{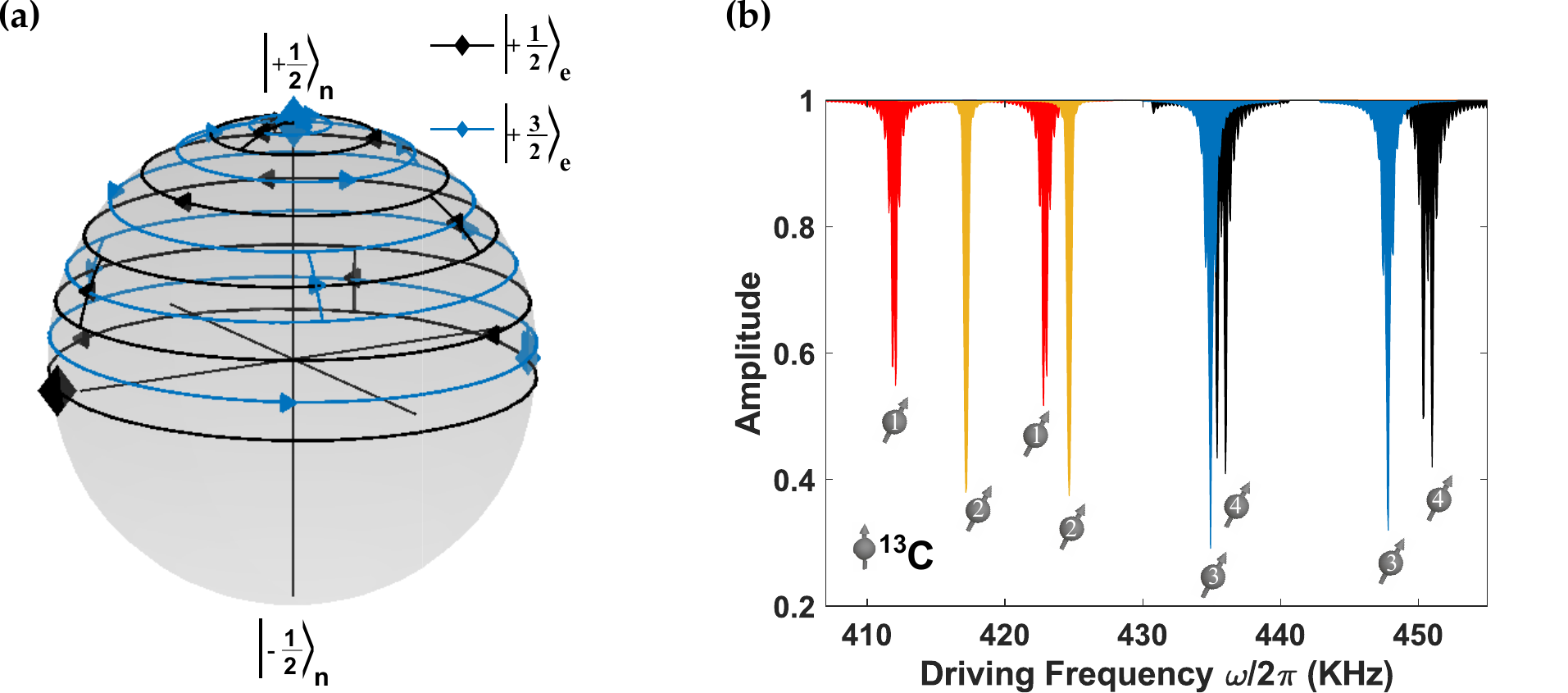}
    \caption{{An example of a $DDrf$ pulse sequence is represented for $N=3$. Henceforth there are $N+1=4$ $rf$ nuclear spin driving pulses in the sequence}}
    \label{fig6part2}
\end{figure}

{The $rf$ pulse sequence as described in Fig.  2(a) of main text is used for rotation of nuclear spin depending on the initial spin state of the $V_{Si}^-$-center electron spin. The $\pi$ pulses which are represented by yellow blocks would flip the $V_{Si}^-$-center electron spin $\ket{\frac{3}{2}}_e\Leftrightarrow\ket{\frac{1}{2}}_e$. These $\pi$ pulses acts as a high pass filter, henceforth preserving the coherence of the $V_{Si}^-$-center. Such a sequence that preserves the $V_{Si}^-$-center electron spin coherence through $\pi$ pulses are known as a dynamic decoupling (DD) sequences. Since this sequence is initerleaved with the $rf$ pulses, the complete sequence is called $DDrf$ sequence. Hence if the initial spin state of the color center is $\ket{\frac{3}{2}}_e$, the initial $rf$ pulse  with frequency $\omega = \omega_1$  in $DDrf$ sequence would drive the nuclear spin between $\ket{\frac{1}{2}}_n\Leftrightarrow\ket{-\frac{1}{2}}_n$. The direction of driving the nuclear spin  on bloch sphere is controlled by the phase of the $rf$ pulse. After the spin flip of color center spin to $\ket{\frac{1}{2}}_e$ the $rf$ pulse would not interact with nuclear spin, hence leading to free evolution around z-axis of bloch sphere. Hence this way the rotation of nuclear spin is dictated by the initial state of the color center spin. An example of $DDrf$ pulse sequence using three $\pi$ pulses is depicted in Fig. \ref{fig6part2}}

Assuming $N$ number of $\pi$ pulses with an initial spin state of the $V_{Si}^-$-center being $\ket{m_s}_e$, the nuclear spin state evolution can be described through $N+1$ unitary matrices $V_{m_s}^k$ where $k=1,2,...N+1$. The first unitary matrix [Eq. (\ref{eqn12})] defines the driving of nuclear spin using the first $rf$ pulse  for a period of $\tau_n$. The frame of reference has to be shifted after the $\pi$ pulse owing to the spin flip of $V_{Si}^-$-center electron spin. Hence the transpose of rotation matrix $R_{m_s}^T (\tau_n) = e^{(-i\omega \tau_n I_{z,\beta_{m_s}})}$ is performed.
\begin{equation}
\begin{split}
V_{m_s}^1 = R_{m_s}^T (\tau_n) U_{m_s}(\tau_n,\omega,\phi_{1})
\label{eqn12}
\end{split}
\end{equation}
 The consecutive unitary matrices consist of a rotation matrix to shift the frame of reference corresponding to the $V_{Si}^-$-center electron spin state, a unitary matrix corresponding to nuclear spin driving for a period of $2\tau_n$ and finally a transposed rotation matrix as depicted in  Eq. (\ref{eqn13}). The unitary matrix description depends on the count of the rf pulse $k$ and on the spin state of the  $V_{Si}^-$-center electron spin $m_s$.


\begin{equation}
V_{m_s}^{k>1}= 
\begin{cases}
      R_{3/2}^T ((2k-1)\tau_n) U_{3/2}(2\tau_n,\omega,\phi_{k}) R_{3/2} ((2k-3)\tau_n) & m_s =1/2  , \, k\,even\\
      R_{1/2}^T ((2k-1)\tau_n) U_{1/2}(2\tau_n,\omega,\phi_{k}) R_{1/2} ((2k-3)\tau_n) & m_s =1/2  , \, k\,odd\\
      R_{1/2}^T ((2k-1)\tau_n) U_{1/2}(2\tau_n,\omega,\phi_{k}) R_{1/2} ((2k-3)\tau_n) & m_s = 3/2  , \, k\,even\\
      R_{3/2}^T ((2k-1)\tau_n) U_{3/2}(2\tau_n,\omega,\phi_{k}) R_{3/2} ((2k-3)\tau_n) & m_s =3/2  , \, k\,odd\\
\end{cases}
\label{eqn13}
\end{equation}

The final unitary matrix as depicted in  Eq. (\ref{eqn14}) contains a rotation matrix to shift the frame of reference corresponding to the $V_{Si}^-$-center electron spin state and the driving of nuclear spin for a period of $\tau_n$.  Upon defining the unitary matrices for each section of $rf$ sequence the overall unitary matrix ($\Tilde{V}_{m_s}$) is matrix multiplication of all the unitary matrices in chronological order as shown in Eq. (\ref{eqn14}).
\begin{equation}
V_{m_s}^{N+1}= 
\begin{cases}
      U_{3/2}(\tau_n,\omega,\phi_{N+1}) R_{3/2} ((2N-1)\tau_n) & m_s =1/2  , \, N\,odd\\
      U_{1/2}(\tau_n,\omega,\phi_{N+1}) R_{1/2} ((2N-1)\tau_n) & m_s =1/2  , \, N\,even\\
      U_{1/2}(\tau_n,\omega,\phi_{N+1}) R_{1/2} ((2N-1)\tau_n) & m_s = 3/2  , \, N\,odd\\
      U_{3/2}(\tau_n,\omega,\phi_{N+1}) R_{3/2} ((2N-1)\tau_n) & m_s =3/2  , \, N\,even\\
\end{cases}
\label{eqn14}
\end{equation}

\begin{equation}
\Tilde{V}_{m_s} = \prod_{k = N+1}^{1} V_{m_s}^{k}
\label{eqn15}
\end{equation}

The phase with which the nuclear spins are driven play a vital role in achieving a controlled rotation. In order to achieve a controlled rotation it is necessary to make sure that the phase of adjacent $rf$ pulses are shifted by $\pi$. The phase of the pulse should account for the period of free evolution as well. Hence considering these criterias the phase of the consecutive $rf$ pulses are given by Eq. (\ref{eqn16}). An example of nuclear spin evolution from $\ket{1/2}_n$ is shown in Fig. \ref{fig1Part2}(a). {The bloch sphere representation exhibits the alternating driving of nuclear spin and free evolution about z-axis coming due to the $DDrf$ sequence}   

\begin{equation}
\begin{split}
\phi_{\tau_n} = (\omega_2-\omega_1)\tau_n\\
\phi_{1} = \phi_{initial}\\
\phi_{2} = \phi_1 + \phi_{\tau_n} + \pi\\
\phi_{l+2} = \phi_{l} + 2\phi_{\tau_n}\\
\phi_{N+1} = \phi_{N-1} + \phi_{\tau_n}\\
\label{eqn16}
\end{split}
\end{equation}
Upon performing frequency sweep of the pulse sequence from Fig. \ref{fig1Part2}(b) a spectrum from the amplitude contrast can be obtained as shown in Fig. \ref{fig1Part2}(b). Each nuclear spin identified from the spectrum has 2 dips which corresponds to the $\omega=\omega_1$ and $\omega=\omega_2$. 

\begin{equation}
\begin{split}
V = \ket{3/2}_{ee}\bra{3/2}\otimes \Tilde{V}_{3/2} +\\
\ket{1/2}_{ee}\bra{1/2}\otimes \Tilde{V}_{1/2}
\label{eqn17}
\end{split}
\end{equation}

\subsection{Independent driving}
The nuclear spins that are present within the sensing volume shown in Figs. 2 and 3(a) are identifiable individually. However with a multiple number of nuclear spin within the sensing volume there is a possibility that a single frequency will be able to drive multiple nuclear spins at once. Hence, this makes it challenging to perform nuclear spin manipulation or readout. Henceforth, independent driving of nuclear spin in an environment filled with other nuclear spin is necessary for controllable access. The influence of non independent driving on isotopic concentration is dependent on the choice of nuclear spin driving time.  
\begin{figure}[t]
    \centering
    \includegraphics[width=\textwidth]{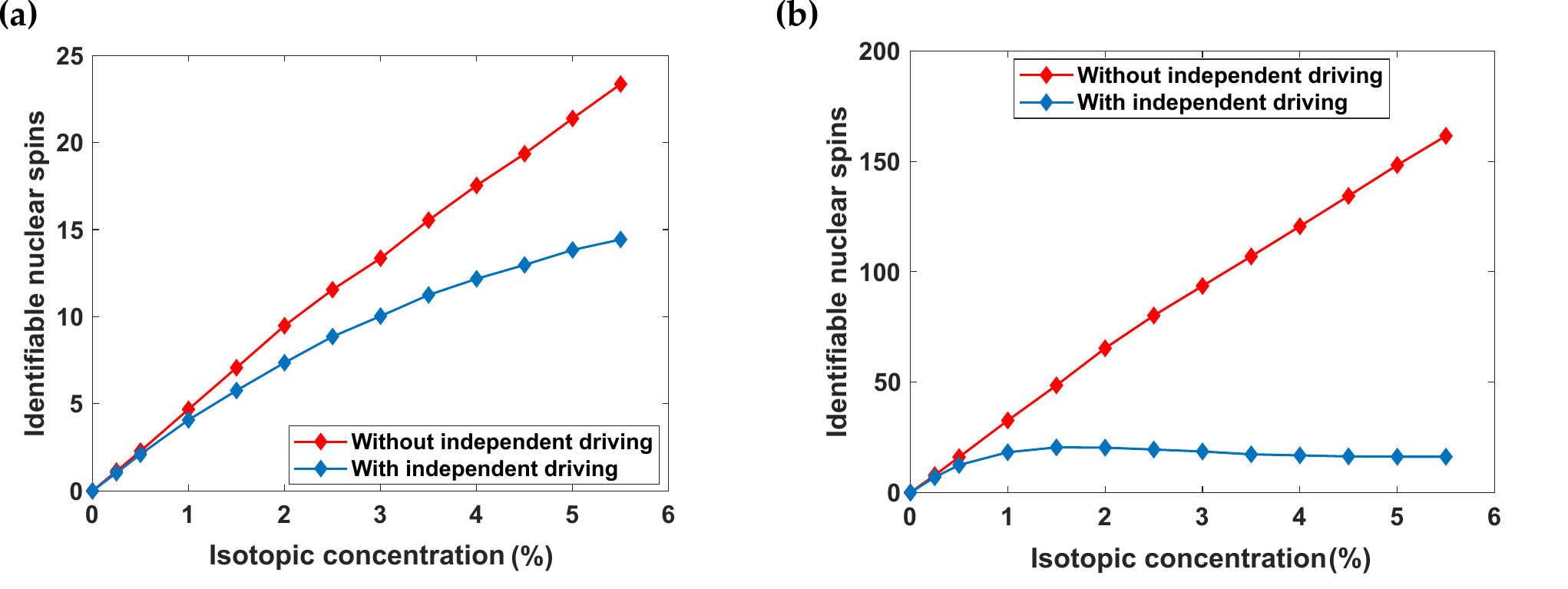}
    \caption{The independent driving is a pivotal criteria for isotope nuclear spin control. The number of nuclear spins that are identified with and without independent driving is represented in (a) and (b) for $\tau_n = 46$ $\mu s$ with $N=20$ and for $\tau_n = 93$ $\mu s$ with $N=100$ respectively.}
    \label{fig8}
\end{figure}
A small sensing volume for a shorter driving time of $\tau_n=46$ $\mu s$ with  $N=20$ indicates that the dominance of non independent driving is noticed only for a larger concentration of isotope nuclear spins as shown in Fig. \ref{fig8}(a). A larger sensing volume for $\tau_n=93$ $\mu s$ with  $N=100$ as depicted in Fig. 2 implies that the dominance of non independent driving increases as the isotopic concentration increases [Fig. \ref{fig8}(b)].  
\section{Spin Coherence}

\subsection{CPMG}
\begin{figure}[tbh]
    \centering
    \includegraphics[width=\textwidth]{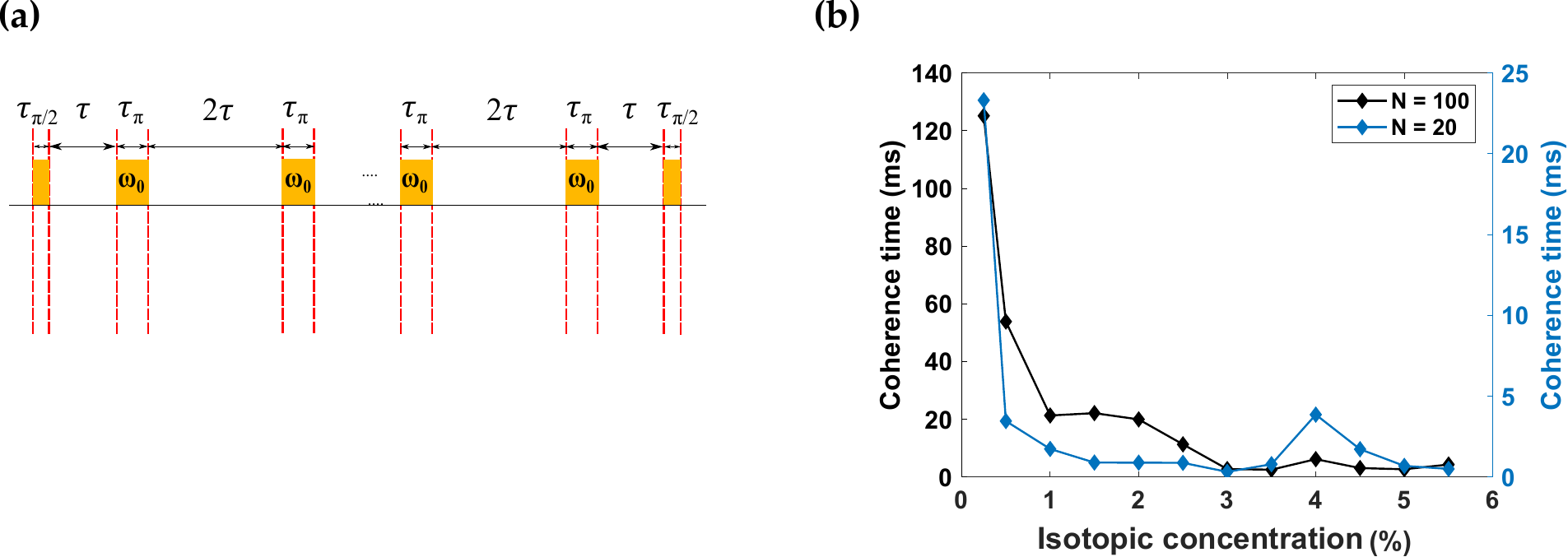}
    \caption{(a)The sequence for examining the coherence of the $V_{Si}^-$-center (CPMG) is depicted [S4]. This sequence would be helpful in keeping the electron spin coherence alive for a longer time using the $\pi$ pulses. (b) The coherence analysis with varying isotope concentration reveals the deteriorating coherence time with increasing concentration hence justifying the need for isotopic purity.}
    \label{fig9}
\end{figure}
The coherence property of the $V_{Si}^-$-center electron spin is dependent on  the nuclear spin bath. In order to examine the coherence property of the color center, a dynamic decoupling sequence namely CPMG (Carr-Purcell-Meiboom-Gill) sequence is employed as shown in Fig. \ref{fig9}(a).

The coherence analysis is carried out through the method of cluster correlation expansion (CCE) [26,28]. In an environment filled with $K$ nuclear spins and a $V_{Si}^-$-center the source of decoherence is usually the magnetic dipole-dipole interaction as shown in Eq. (\ref{eqn18}) as $\Tilde{A_{lk}}$. The indices $l$ and $k$ represents the spin entity within the environment. The  $V_{Si}^-$-center is given an index of 0 while the isotope nuclear spin are indexed with natural numbers. 

The unitary matrix that represents the bifurcated evolution of the nuclear spins in the nuclear spin bath is described in Eq. (\ref{eqn19}). 
\begin{equation}
\begin{split}
U_{m_s}^{l,k}(t) = e^{(-iH_{m_s}^{l,k}t)}
\label{eqn19}
\end{split}
\end{equation}

The spin state of the nuclear spins and nuclear spin pairs in the bath is given by Eq. (\ref{eqn20}).  
\begin{equation}
\begin{split}
\ket{\psi_{0,l}} = \ket{\psi_l}\\
\ket{\psi_{l,k}} = \ket{\psi_l}\otimes\ket{\psi_k}
\label{eqn20}
\end{split}
\end{equation}

The unitary matrix that describes the spin state evolution for initial $V_{Si}^-$-center electron spin state as $m_s = 3/2$ and $m_s = 1/2$ is depicted as Eq. (\ref{eqn21}). 

\begin{equation}
\begin{split}
V_{3/2}^{l,k}(\tau) = U_{3/2}^{l,k}(\tau)U_{1/2}^{l,k}(2\tau)U_{3/2}^{l,k}(2\tau).....U_{1/2}^{l,k}(2\tau)U_{3/2}^{l,k}(\tau)\\
V_{1/2}^{l,k}(\tau) = U_{1/2}^{l,k}(\tau)U_{3/2}^{l,k}(2\tau)U_{1/2}^{l,k}(2\tau).....U_{3/2}^{l,k}(2\tau)U_{1/2}^{l,k}(\tau)
\label{eqn21}
\end{split}
\end{equation}
For a pure spin bath, the correlation function for a single nuclear spin $l$ is given by Eq. (\ref{eqn22}). Similarly, the correlation function for a nuclear spin pair $\{l,k\}$ is given by Eq. (\ref{eqn23}).
\begin{equation}
\begin{split}
\mathfrak{L}_{\{l\}}(\tau) = \bra{\psi_{l}}(V_{1/2}^{0,l}(\tau))^T V_{3/2}^{0,l}(\tau)\ket{\psi_{l}}\\
\label{eqn22}
\end{split}
\end{equation}

\begin{equation}
\begin{split}
\mathfrak{L}_{\{l,k\}}(\tau) = \frac{\bra{\psi_{l,k}}(V_{1/2}^{l,k}(\tau))^T V_{3/2}^{l,k}(\tau)\ket{\psi_{l,k}}}{\mathfrak{L}_{\{l\}}\mathfrak{L}_{\{k\}}}
\label{eqn23}
\end{split}
\end{equation}
Cluster correlation function for single nuclear spins and nuclear spin pairs are represented in Eqs. (\ref{eqn24}) and (\ref{eqn25}) respectively. Hence the overall  correlation function $\Tilde{\mathfrak{L}(\tau)}$ for CCE-2 is Eq. (\ref{eqn26}). A CCE-N  method considers interaction of $V_{Si}^-$-center with a cluster of $N$ nuclear spins. The coherence function resembles a stretched exponential with noise coming from nuclear spin interaction. The decay time of the stretched exponential yields the coherence time of the sample.  
\begin{equation}
\begin{split}
\Tilde{\mathfrak{L}^1}(\tau) = \prod_{l=1}^K \mathfrak{L}_{\{l\}}(\tau)\\
\label{eqn24}
\end{split}
\end{equation}

\begin{equation}
\begin{split}
\Tilde{\mathfrak{L}^2}(\tau) = \prod_{l,k=1}^K \mathfrak{L}_{\{l,k\}}(\tau)\\
\label{eqn25}
\end{split}
\end{equation}

\begin{equation}
\begin{split}
\Tilde{\mathfrak{L}(\tau)} =  \Tilde{\mathfrak{L}^1(\tau)}\Tilde{\mathfrak{L}^2(\tau)}\\
\label{eqn26}
\end{split}
\end{equation}

\subsection{Variation with concentration}

Using the analysis from the above section an analysis was conducted to examine the effect of isotopic concentration on the coherence time under two scenarios where the number of $\pi$ pulses ($N$) is varied (i) $N=100$ and (ii) $N=20$. The median analysis for each concentration is carried out for 5 random distributions of isotope nuclear spin around the $V_{Si}^-$-center. For each distribution a statistical average of 5 random spin configuration in the nuclear spin bath is considered. The decreasing coherence time dependence on increasing isotopic concentration is depicted in Fig. \ref{fig9}(b). The coherence analysis hence justifies the necessity of isotopic purity in our sample in order to preserve the $V_{Si}^-$-center electron spin coherence for a longer time.

\newpage
\renewcommand*\labelenumi{[S\theenumi]}

\begin{enumerate}
    \item Kobayashi, Takuma et al. (2019). "Native point defects and carbon clusters in 4H-SiC: A hybrid functional study" \textit{Journal of Applied Physics } 125, 125701 (2019); https://doi.org/10.1063/1.5089174
    \item L.-P. Yang, C. Burk, M. Widmann, S.-Y. Lee, J. Wrachtrup, N. Zhao, "Electron spin decoherence in silicon carbide nuclear spin bath." \textit{Phys. Rev. B} 90, 241203(2014)
    \item C.E. Bradley, J. Randall, M.H. Abobeih, R.C. Berrevoets, M.J. Degen, M.A. Bakker, M. Markham, D.J. Twitchen, T.H. Taminiau, A Ten-Qubit Solid-State Spin Register with Quantum Memory up to One Minute. \textit{Phys. Rev. X} 9, 031045 (2019)
    \item Ma, WL., Wolfowicz, G., Zhao, N. et al. Uncovering many-body correlations in nanoscale nuclear spin baths by central spin decoherence. \textit{Nat Commun} 5, 4822 (2014). https://doi.org/10.1038/ncomms5822
\end{enumerate}

\end{document}